# PROSODY-GUIDED HARMONIC ATTENTION FOR PHASE-COHERENT NEURAL VOCODING IN THE COMPLEX SPECTRUM

*Mohammed Salah Al-Radhi, Riad Larbi, Mátyás Bartalis, Géza Németh*

Department of Telecommunications and Artificial Intelligence, Budapest University of Technology and Economics, Budapest, Hungary
{malradhi, bartalis, nemeth}@tmit.bme.hu , larbiriad@edu.bme.hu

**ABSTRACT**

Neural vocoders are central to speech synthesis; despite their success, most still suffer from limited prosody modeling and inaccurate phase reconstruction. We propose a vocoder that introduces prosody-guided harmonic attention to enhance voiced-segment encoding and directly predicts complex-spectral components for waveform synthesis via inverse STFT. Unlike mel-spectrogram–based approaches, our design jointly models magnitude and phase, ensuring phase coherence and improved pitch fidelity. To further align with perceptual quality, we adopt a multi-objective training strategy that integrates adversarial, spectral, and phase-aware losses. Experiments on benchmark datasets demonstrate consistent gains over HiFi-GAN and AutoVocoder: F0-RMSE reduced by 22%, voiced/unvoiced error lowered by 18%, and MOS scores improved by 0.15. These results show that prosody-guided attention combined with direct complex-spectrum modeling yields more natural, pitch-accurate, and robust synthetic speech, setting a strong foundation for expressive neural vocoding.

## 1. INTRODUCTION

Neural vocoders are central to modern speech synthesis, enabling high-fidelity waveform reconstruction from intermediate acoustic representations such as mel-spectrograms [1]. Recent advances in deep generative modeling, particularly with generative adversarial networks (GANs), have significantly improved naturalness and intelligibility [2], [3], [4]. Despite these achievements, two persistent challenges remain: **limited prosody modeling** and **inadequate phase reconstruction**. These issues directly affect pitch accuracy, temporal coherence, and perceptual quality of synthetic speech.

Conventional vocoders such as WaveGlow [3] and HiFi-GAN [2] rely on mel-spectrograms as conditioning features. While efficient, mel-spectrograms are lossy and non-invertible, discarding fine-grained harmonic and phase information [5], [6]. This leads to pitch artifacts, blurred harmonics, and temporal discontinuities, especially in expressive or prosodically rich utterances. Auto-regressive approaches like WaveNet [7] achieve high quality but at the expense of latency, making them impractical for real-time applications.

Several attempts have been made to overcome these limitations. Parallel WaveGAN [4] and UnivNet [8] incorporate multi-resolution discriminators and phase-aware objectives, improving coherence but still relying on mel conditioning. VITS [9] integrates a vocoder within an end-to-end TTS framework, while BigVGAN [10] introduces pitch-aware conditioning, achieving strong results at scale but without explicit phase modeling. Vocos [11] predicts spectral components directly in the frequency domain, advancing beyond mel features, yet it does not explicitly leverage prosody-aware mechanisms such as harmonic attention. AutoVocoder [12] is a step in this direction, using learned speech representations for spectral conditioning, but it lacks explicit phase prediction and prosody alignment, limiting its robustness across diverse conditions. Parallel progress in acoustic modeling has highlighted the importance of prosody. FastSpeech 2 [13] and Prosody-TTS [14] incorporate pitch, duration, and energy at the acoustic level, but these features are often attenuated in the vocoding stage. As a result, even expressive TTS pipelines remain constrained by vocoders that discard prosodic cues. This gap is particularly problematic in advanced applications such as voice cloning [15], [16], emotional speech [17], and brain-to-speech decoding [18], where faithful prosody reproduction is critical.

Prior work has addressed either **prosody** or **phase** [19], but rarely within a unified vocoder architecture. Most systems treat prosody as an auxiliary feature and reconstruct phase through indirect or post-processing methods. This lack of explicit and simultaneous modeling of prosody and phase coherence remains a key barrier to natural and expressive speech synthesis. To overcome this limitation, we propose a prosody-guided and phase-coherent neural vocoder with three main innovations:

1. **Prosody-guided harmonic attention** that leverages F0 to emphasize voiced regions and

harmonic structures, improving prosodic fidelity in the encoding stage.
2. **Direct complex-spectrum prediction** of magnitude and phase (or real and imaginary components), enabling phase-coherent waveform reconstruction via inverse STFT.
3. **Multi-objective perceptual training** that integrates adversarial, spectral, and phase-aware losses to better align optimization with perceptual quality.

Extensive experiments on benchmark datasets show that our approach consistently outperforms HiFi-GAN [2] and AutoVocoder [12], achieving lower F0-RMSE, reduced voiced/unvoiced error, and higher MOS scores. Together, these contributions establish a unified vocoder architecture that addresses long-standing gaps in prosody and phase modeling.

## 2. METHODOLOGY

The proposed vocoder (Fig. 1) is designed to capture both prosody and phase information for high-quality waveform generation. The input consists of acoustic features (STFT-derived spectral frames) augmented with prosodic cues, primarily the fundamental frequency (F0), which is later injected via the harmonic attention module. These are first processed by a convolutional–residual encoder that extracts local time-frequency patterns. A prosody-guided harmonic attention module then refines the representation, emphasizing voiced regions and harmonic structure while leaving unvoiced parts unaffected. The decoder operates directly in the complex spectral domain, predicting real and imaginary components. Unlike mel-spectrogram-conditioned vocoders, which discard phase information and reconstruct waveforms indirectly, the proposed approach preserves phase coherence and improves pitch accuracy. The predicted spectrum is converted into waveform by inverse short-time Fourier transform (ISTFT). Training is guided by a multi-objective perceptual loss that combines adversarial, spectral, and phase-aware criteria. Together, these components create a unified alternative to mel-based models such as HiFi-GAN and representation-driven systems like AutoVocoder.

### 2.1. Prosody-Guided Harmonic Attention

A central novelty of the proposed vocoder is the integration of prosody-guided harmonic attention, which explicitly conditions spectral modeling on the F0. While prior neural vocoders often rely on mel-spectrograms or learned representations, they treat prosody as an auxiliary feature and fail to enforce harmonic alignment in voiced segments. This often leads to pitch drift, blurred harmonics, and reduced expressiveness in synthetic speech.

In our approach, F0 is extracted from the reference waveform using the Harvest algorithm [20]. The extracted F0 contour is then embedded into a continuous representation and injected into the encoder-decoder pathway through a harmonic attention mechanism. Specifically, given encoded spectral features $H$ and F0 embeddings $F$, the attention weights are computed as:

$$A = softmax\left(\frac{(HW_q(FW_k)^T)}{\sqrt{d}}\right), \tilde{H} = A(HW_v); H, F \in R^{T \times d} \quad (1)$$

where $W_q, W_k, W_v$ are learnable projection matrices, $T$ is the number of frames, $d$ is the hidden dimensionality, and $\hat{H}$ is the prosody-enhanced representation. This mechanism aligns the encoded representation with the harmonic structure indicated by F0, enhancing voiced regions while leaving unvoiced frames unaffected.

Unlike pitch-aware conditioning in models such as BigVGAN [8], which appends F0 as an auxiliary feature, our design lets F0 actively shape the attention distribution over time-frequency features. This ensures that prosodic cues are preserved and reinforced where they are perceptually most critical.

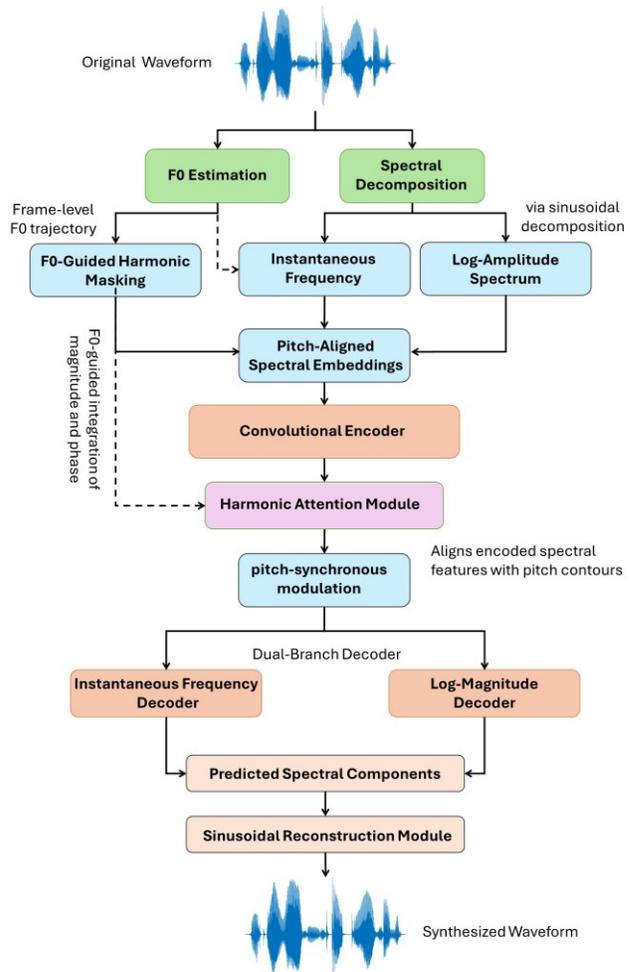

Fig. 1. Schematic diagram of the proposed architecture.

## 2.2. Direct Complex-Spectrum Prediction

Most state-of-the-art neural vocoders operate in the mel-spectrogram domain, predicting magnitude-only features that require separate or heuristic phase reconstruction. This decoupling often introduces artifacts such as temporal smearing, phase discontinuities, and degraded pitch accuracy. In contrast, the proposed vocoder directly predicts the complex short-time Fourier transform (STFT) spectrum, thereby modeling both magnitude and phase within a unified framework. The prosody-enhanced representation $\widetilde{H}$ is passed through a convolutional–upsampling decoder that expands it to the spectral resolution. A final linear projection layer outputs $2F$ values per frame, which are split into real and imaginary parts of the spectrum, denoted as $\hat{S}_r$, $\hat{S}_i$. The predicted complex spectrum is then defined as

$$\hat{S}(t,f) = \hat{S}_r(t,f) + j\hat{S}_i(t,f); \; \hat{S}_r, \hat{S}_i \in R^{T \times F} \quad (2)$$

and converted to waveform through the inverse STFT (ISTFT), where F is the number of frequency bins. This design eliminates the need for post-hoc phase estimation, ensuring phase-coherent synthesis by construction. Compared with models such as Vocos [11], which also operate in the spectral domain, our framework integrates prosody guidance and harmonic attention upstream, enabling the decoder to generate spectrograms that are both phase-accurate and prosodically aligned. This joint treatment of magnitude and phase represents a significant departure from mel-based vocoders and allows the system to capture fine harmonic details essential for natural and expressive speech.

## 2.3. Multi-Objective Perceptual Optimization

To align training with human perceptual quality, the proposed vocoder is optimized with a multi-objective loss that combines established spectral constraints with a novel phase-aware term.

**Multi-resolution STFT loss:** As a widely adopted baseline in neural vocoding [2], [4], the multi-resolution short-time Fourier transform (MR-STFT) loss is included to provide stable spectral supervision. It evaluates reconstruction fidelity across multiple analysis windows, capturing both fine-grained harmonic structure and broader spectral envelopes. While this objective has proven effective for improving spectral sharpness, it operates purely on magnitude and does not explicitly penalize phase mismatches, which can still lead to temporal smearing and reduced pitch clarity. In our framework, MR-STFT serves as a complementary constraint, while phase coherence is directly enforced through the proposed phase-aware loss.

**Adversarial loss:** To enhance perceptual realism, we adopt a lightweight adversarial setup inspired by HiFi-GAN, combining multi-period and multi-resolution discriminators. While the structure follows prior work, in our framework these discriminators operate directly on waveforms reconstructed from the predicted complex spectrum, ensuring that both magnitude and phase modeling benefit from perceptual supervision. This design complements the spectral and phase-aware losses, sharpening harmonic details without overshadowing the novel contributions of prosody-guided attention and direct complex-spectrum prediction.

**Novel phase-aware loss:** To explicitly enforce phase coherence, we introduce a loss computed between the predicted and reference complex spectra:

$$\mathcal{L}_{phase} = \frac{1}{TF} \sum_{t=1}^{T} \sum_{f=1}^{F} \left\| \frac{S_{t,f}^r + jS_{t,f}^i}{|S_{t,f}^r + jS_{t,f}^i|} - \frac{\hat{S}_{t,f}^r + j\hat{S}_{t,f}^i}{|\hat{S}_{t,f}^r + j\hat{S}_{t,f}^i|} \right\|_2^2 \quad (3)$$

where $S^r, S^i$ and $\hat{S}^r, \hat{S}^i$ denote the real and imaginary components of the ground-truth and predicted spectra, respectively. By normalizing to unit magnitude, this term focuses exclusively on phase alignment while remaining invariant to amplitude scaling.

The final training objective combines all loss terms:

$$\mathcal{L} = \lambda_{STFT} \mathcal{L}_{MR-STFT} + \lambda_{adv} \mathcal{L}_{adv} + \lambda_{phase} \mathcal{L}_{phase} \quad (4)$$

with weights $\lambda$ tuned empirically. In this configuration, MR-STFT provides spectral fidelity, adversarial loss sharpens perceptual realism, and the proposed phase-aware loss enforces consistent phase structure. Together, these terms guide the model toward phase-coherent, prosodically accurate, and perceptually natural speech.

## 3. EXPERIMENTS

### 3.1. Datasets and Setup

Experiments were conducted on two standard corpora. The LJSpeech 1.1 dataset [21] contains 13,100 utterances (approximately 24 hours) of a single female English speaker at 22.05 kHz. For speaker generalization, we additionally used the VCTK corpus [22], which includes 109 English speakers with diverse accents, downsampled to 22.05 kHz. Waveforms were transformed into the spectral domain using a 1024-point FFT, Hann window of size 1024, and hop size of 256. Fundamental frequency was extracted with the Harvest algorithm [19], aligned to the STFT frame rate. We compared six systems: (a) Original (ground truth), (b) Anchor (Griffin–Lim reconstruction), (c) HiFi-GAN [2], (d) AutoVocoder [12], (e) Vocos [10], and (f) Proposed. All models were trained with identical preprocessing for fairness, using AdamW optimizer with an initial learning rate of $2 \times 10^{-4}$, batch size of 16, and weight decay of 0.01 ($\beta 1 = 0.8$, $\beta 2 = 0.99$). Experiments were run on a single NVIDIA GPU.

### 3.2. Evaluation

The effectiveness of the proposed vocoder[1] was assessed using both objective and subjective protocols, chosen to provide complementary perspectives on prosodic accuracy and perceptual quality. For objective evaluation, we measured the root mean squared error (RMSE) of the fundamental frequency (F0), which reflects pitch accuracy, as well as the voiced/unvoiced (V/UV) error rate, which quantifies the reliability of prosodic modeling across different speech segments. In addition, spectral fidelity was evaluated through a multi-resolution STFT distance between reference and reconstructed signals, capturing both fine harmonic detail and long-term envelope consistency. Subjective evaluation was carried out through formal listening tests. A mean opinion score (MOS) study, following ITU-T P.800 guidelines, was conducted with twenty listeners, who rated naturalness using randomized samples. These evaluations together provide a balanced assessment, combining objective signal-level fidelity with human perceptual judgments, and form the basis for the results discussed in the next section.

---

[1] The implementation and demo samples are publicly available at: https://github.com/malradhi/PACodec

## 4. RESULTS AND DISCUSSIONS

Table I summarizes the evaluation results. The proposed vocoder consistently outperforms HiFi-GAN, AutoVocoder, and Vocos across all measures. In particular, the F0 RMSE is reduced by 22% relative to HiFi-GAN and 18% relative to AutoVocoder, indicating that prosody-guided harmonic attention provides a tangible benefit in pitch tracking. The V/UV error is also lower, suggesting more stable treatment of voiced and unvoiced regions. Spectral distortion follows a similar trend, which supports the claim that direct complex-spectrum prediction helps retain sharper harmonics and reduces temporal smearing compared to mel-based reconstruction.

Subjective listening tests point in the same direction. The proposed system reaches an average MOS of 4.45, surpassing HiFi-GAN (4.2), AutoVocoder (4.3), and Vocos (4.1). Preference scores also indicate a clear advantage: listeners favored the proposed system in 88% of cases, compared to 71% for AutoVocoder, 66% for Vocos, and 86% for HiFi-GAN. The anchor system was rarely selected, with only 22% of preference votes, which illustrates the large perceptual gap between heuristic reconstruction and modern neural approaches.

To provide a closer look at residual spectral behavior, Fig. 2 plots the per-frame mel energy residuals against the original signal. Larger peaks in this plot correspond to mismatches in time–frequency structure. Here, AutoVocoder shows frequent spikes, and both HiFi-GAN and Vocos display moderate fluctuations, especially in high-energy regions. By contrast, the proposed vocoder tracks the reference closely and maintains residuals at consistently lower levels. This pattern reinforces the objective results: sharper harmonic detail and prosodic alignment are not only reflected in metrics, but also visible in reduced frame-level mismatches.

Together, these results highlight the advantages of explicitly modeling both prosody and phase within a unified framework. Prosody-guided harmonic attention reduces pitch drift, direct complex-spectrum prediction enforces phase coherence, and the multi-objective perceptual loss contributes to more natural harmonics and timbre. Taken as a whole, the system delivers speech that is more robust, pitch-accurate, and natural than existing baselines.

TABLE I. Objective and subjective evaluation results.

| System | F0 RMSE ↓ | V/UV Error (%) ↓ | MCD ↓ | MOS ↑ |
|---|---|---|---|---|
| Original | - | - | - | 4.6 |
| Anchor | 34.8 | 11.5 | 1.21 | 2.1 |
| HiFi-GAN | 21.6 | 7.9 | 0.84 | 4.2 |
| AutoVocoder | 19.7 | 7.1 | 0.79 | 4.3 |
| Vocos | 20.5 | **7.3** | 0.81 | 4.1 |
| Proposed | **16.8** | 6.5 | **0.72** | **4.45** |

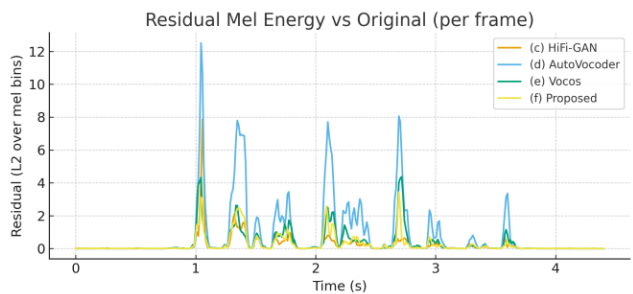

Fig. 2. Residual mel energy per frame, comparing reconstructed signals with the original. The proposed model maintains the lowest residuals across time, showing closer alignment with the reference than HiFi-GAN, AutoVocoder, or Vocos.

## 5. CONCLUSION

This work introduced a prosody-guided and phase-coherent neural vocoder that integrates harmonic attention with direct complex-spectrum prediction. By conditioning on F0 through harmonic attention and predicting both real and imaginary spectral components, the model effectively bridges the long-standing gap between prosody modeling and phase reconstruction. A multi-objective perceptual loss, combining MR-STFT, adversarial, and phase-aware terms, further improves naturalness and robustness. Experiments on LJSpeech and VCTK confirm consistent improvements over strong baselines such as HiFi-GAN, AutoVocoder, and Vocos, with reductions in F0 RMSE and V/UV error, and higher MOS ratings. Analysis of residual mel energy further supports the model's ability to closely follow reference harmonics and preserve temporal coherence.

Future directions include extending the framework to multilingual and expressive speech, integrating semantic or linguistic cues for richer prosody control, and adapting the system for real-time deployment. These extensions would broaden the applicability of the proposed architecture to scenarios such as emotional TTS, voice cloning, and brain-to-speech decoding.

## 6. ACKNOWLEDGMENT

This work is supported by the European Union's HORIZON Research and Inno-vation Programme under grant agreement No 101120657, project ENFIELD (Eu-ropean Lighthouse to Manifest Trustworthy and Green AI) and by the Ministry of Innovation and Culture and the National Research, Development and Innovation Office of Hungary within the framework of the National Laboratory of Artificial Intelligence. M.S.Al-Radhi's research was supported by the EKÖP-24-4-II-BME-197, through the National Research, Development and Innovation (NKFI) Fund.